\documentclass{article}[10pt]
\def\case#1/#2{\textstyle\frac{#1}{#2}}
\usepackage{graphicx}
\begin{document}
\title{The Evolution of Our Local Cosmic Domain:\\ Effective Causal Limits}
\author{G. F. R. Ellis$^1$ \and W. R. Stoeger$^{1,2}$}

\maketitle

$^1$ Mathematics Department, University of Cape Town, Rondebosch, Cape Town 7701\\

$^2$ Vatican Observatory Research Group, Steward Observatory, The
University of Arizona, Tucson, AZ 85721.

\maketitle

\begin{abstract}
The causal limit usually considered in cosmology is the particle
horizon, delimiting the possibilities of causal connection in the
expanding universe. However it is not a realistic indicator of the
effective local limits of important interactions in spacetime. We consider here the
\emph{matter horizon} for the Solar System, that is, the comoving
region which has contributed matter to our local physical
environment. This lies inside the \emph{effective domain of
dependence}, which (assuming the universe is dominated by dark matter
along with baryonic matter and vacuum-energy-like dark energy)
consists of those regions that have had a
significant active physical influence on this environment through effects
such as matter accretion and acoustic waves. 
It is not determined by the velocity of light $c$, but by the flow of matter perturbations
along their world lines and associated gravitational effects. We emphasize how small a region the
perturbations which became our Galaxy occupied, relative to the
observable universe -- even relative to the smallest- scale
perturbations detectable in the cosmic microwave background
radiation. Finally, looking to the future of our local cosmic
domain, we suggest simple dynamical criteria for determining the
\emph{present domain of influence} and the \emph{future matter
horizon}. The former is the radial distance at which our local
region is just now separating from the cosmic expansion. The latter
represents the limits of growth of the matter horizon in the far
future.
\end{abstract}
 
\noindent
{\Large keywords \\
\noindent
cosmology: theory}

\section{INTRODUCTION}

Causal limits in cosmology are usually taken as indicated by the
past light cone. This paper aims to introduce a more nuanced view
of these limits, based on timelike world lines.
 
In probing the universe as a whole -- as a single object of study -- we
peer out to collect and interpret data arriving at the speed
of light from its farthest reaches, and therefore from its earliest
epochs. Much of observational cosmology relies on
precise measurements of the cosmic microwave background radiation
(CMB), which emanates from the last-scattering surface (LSS) where it
last interacted significantly with matter some 13.7 billion years
ago. Because of the opaqueness of the early universe, this is
the most distant region from which we can receive
electromagnetic signals, a  scale of at least $10^{29}$
cm. However, these signals are difficult to detect and have
negligible influence on the universe today; while they represent
absolute limits to observation, they themselves do not exert a significant
influence on present-day conditions in local regions. The same is true
for other particles that may arrive from very distant regions, such as
neutrinos. Even though
the global expanding cosmic environment from Planck-era exit until today
has been important for establishing conditions of large-scale near-homogeneity
and consequent relative dynamical
isolation of local regions, large-scale global influences have not,
since inflation, in any other way actively affected evolution on present
comoving scales larger than about 100 Mpc.
 
This analysis presumes that the
present cosmological picture is more or less correct -- i. e. that the dominant
mass-energy constituents in the universe are cold dark matter and vacuum-like
dark energy, along with baryonic matter and radiation. We shall, as standard, 
refer to this as the $\Lambda$CDM model, and will assume that it, or something not
too much different from it, is a correct description of the universe's principal
mass-energy components. What follows is only true if there are no exotic
components of the universe (dynamical dark energy, cosmic strings, etc.) with
either relativistic peculiar velocities, relativistic speeds of sound, very 
large Compton wavelengths or even more extreme ideas such as variable speed of 
light theories or deviations from Einstein gravity. There is always the possibility
that our lack of understanding about dark matter, dark energy, and the
lower-multipole anomalies in the CMB may be signaling the presence of such 
exotic mass-energy constituents, or a large-scale modification of gravity, which
would alter our assessment. But these and other unsolved cosmological problems
may also turn out to be understood in terms completely consistent with
the causal mass-energy picture we now have, and we will from here on proceed on 
that basis. And on that basis, most of what we can see by astronomical 
observations has negligible effect on us. If one of the more exotic possibilities
were to turn out to be true, what follows would need modification in accord
with those discoveries.

There is therefore some interest in investigating the space-time region that
\emph{does} make a significant difference to local conditions.\footnote{Colloquially:
which part of space-time makes a real difference to you, as you walk down the 
street or drive a car?} We can do so by reversing our perspective and looking carefully at
the cosmological history of our own local domain and the
`geological' evidence it contains about the early universe. We
distil cosmological significance from it, for example, by relating
the current abundances of elements in our neighbourhood to
baryosynthesis and nucleosynthesis close to our world line in the
very distant past, long before the LSS was
established. Our local domain has a history stretching back through
the epochs of recombination, the Hot Big Bang radiation dominated era, and
inflation, to the end of the Planck era and the start of the classical
universe. And it has a future related to nearby galaxies
and clusters of galaxies -- all the large-scale ``perturbations''
with which it will eventually interact as the universe evolves
towards its distant future.
 
How then can we characterize causality related to our local cosmic domain?
Specifically: how large is
the region in the past which significantly contributed to conditions
in our local physical system? How large is the region which
presently interacts with it? How large is the region that will
affect it in the future? Our primary aim is to define precisely
effective domains of dependence, that is, the portion of the initial
data which makes a significant difference to the evolution
of our local systems -- to our Galaxy and our Solar System
-- from some initial time in the past up to the present time. This
depends on the epoch at which we evaluate it. We would like to know
how large it was at recombination, at the end of inflation, and
prior to inflation; and even further back at its very origins.  

\subsection{The effective horizon}
The basic limit of possible causal influence in a space time (the
past light cone) leads to the absolute limit of causal influence
in cosmology as we extend our past light cone to the origin of the
universe. This absolute limit is the \emph{particle horizon},
defined as the limiting sphere of the most distant matter that
could have influenced us up to the present time by any interaction
proceeding at speeds less than or equal to the speed of light
(Rindler 1956, Hawking and Ellis 1973, Tipler, Clarke and Ellis
1980); most studies of causality in cosmology focus on these 
causal limits. However this is not an adequately nuanced representation of the
domain of influence significantly affecting our local cosmic neighborhood,
because, as far as we have been able to determine on the basis of 
the $\Lambda$CDM model, the influences that make a significant difference to 
conditions in the Solar System and our Galaxy (the Milky Way, or
\emph{MW}) are mediated through massive rather than massless
particles; and these particles travel at very low speeds relative
to the cosmic rest frame. This is related to the fact that the
characteristics for scalar perturbations of pressure-free matter
are timelike world lines (Ehlers, Prasanna and Breuer 1987), and
these are far more important in cosmological history (as presently
understood) than vector or tensor perturbations.
 
Thus, in the $\Lambda$CDM universe model, and in our
universe insofar as it is described by that model, effective causal limits 
are based on timelike world lines of the matter and associated local
gravitational fields, rather than on influences travelling on null geodesics.
At any finite time $t$, these worldlines determine the \emph{Effective
Domain of Influence}, which is the limiting sphere of matter that
can have had a significant effect on our local region from time $t$ to
the present time $t_0$, with comoving radius $r_{ed}(t, t_0)$. 
Taking the limit of the time $t$ towards the start of
the universe defines our \emph{Effective Horizon}, 
which is the analogue of the particle horizon: it
is the limit of the spacetime region that contains all matter and
events that have contributed significantly to the specific history and
characteristics of our local domain since the start of the universe. 
It is thus the boundary of the
total set of events that have made a noticeable difference to what
is happening here and now, through the effects of the massive
particles that have in fact determined local conditions on and near the
Earth. Our lives have would have been significantly different if
conditions there had been altered.

\subsection{The matter horizon}
From this perspective our local domain of influence back
to decoupling consists of the matter in density perturbations
which evolved into our Galaxy, including nearby matter perturbations with which they
merged or interacted. If we take our Galaxy as our present local cosmic
domain -- as the smallest object which we can be reasonably sure has an
identifiable interaction history with us without significantly involving larger
scales -- we can consider its history as it passes through the inflationary
process, reheating, the radiation dominated epoch, the radiation-matter
equality transition, etc., going all the way back to when it was a cluster
of quantum fluctuations just after the Big Bang. 
 
This history reveals that the actual
domains of influence affecting local physical systems in the
universe today since recombination at $z_{rec} \approx 1100$, are, for all
practical purposes, confined to the
bundle of world-lines gradually bringing matter together to form our Galaxy and our
Solar System. The significant factor to be considered at times since decoupling 
is the effective local velocities by which matter was added to the Galaxy from
nearby regions. Indeed one can claim, within the present provisionally
accepted $\Lambda$CDM model, that the main locally significant causal domain --
the region from which the vast majority of constituent matter originated
\footnote{Leaving aside neutrinos and other possible weakly interacting matter components,
as well as cosmic rays from distant galaxies, which -- in accord with present
evidence -- we assume have little effect on the evolution of our local domain.} -- was
characterized by the same comoving matter even long before that, being relatively unchanged
since the decoupling of the dominant cold-dark matter (CDM) at
$z_{CDM}$. Its present comoving size is about 2 Mpc, i.e. $6
\times 10^{25}$cm.
 
We call this bounding sphere the \emph{matter horizon}. This horizon, then, is the limit
beyond which no significant matter has been contributed to our local domain today. It obviously evolves
and is growing at present, as it has in the past when matter was
accreted onto our local high-density region. Its starting point is
at the end of inflation, when the high vacuum left by the rapid
inflationary expansion was filled with matter created by decay of
the inflaton field. This is the originating event for the matter
which exists in and on the earth today. The original matter
(quarks and electrons?) has since been processed in many ways, in
particular through baryosynthesis and nucleosynthesis; but these
were local processes that did not significantly alter the domain
bounded by the matter horizon. 
 
At earlier times, before decoupling at $t_{rec}$, acoustic oscillations,
which travel at the velocity of sound $c_s$, established correlations
among density perturbations within a somewhat larger region, and so
allowed a larger domain to influence local densities, even though
no significant exchange of matter took place between the relevant 
regions. That is why the effective domain of influence can
grow to be larger than indicated by the matter horizon, at times before
decoupling.
 
\subsection{The present domain of influence and the future horizon} 
We shall find it natural to define also a \emph{future matter
horizon} and a \emph{present domain of influence}, specifying the
radius from our position beyond which collapse towards us will
never occur in the future, and the radius beyond which collapse is
not occurring now, respectively. It is easy to confirm that these
are also much less than both the particle horizon and
the visual horizon.\footnote{The visual horizon is the limiting
sphere beyond which we cannot now see regions of the universe via
unimpeded electromagnetic signals. It is defined by the matter
world lines extending to the future from the
intersection of our past light cone with the last-scattering
surface (LSS). We have not yet received any direct information via e-m
radiation from world lines beyond these.}  The very fact that
distant galaxies are now expanding away from us -- and most of
them will continue to do so forever -- is secure evidence of this.
 
In what follows, we aim to determine the sizes of these effective domains
for the MW in a physically rigorous and potentially measurable way. Our
discussion can be easily ``scaled up'' to a larger region -- e. g. the 
Local Group -- simply by considering the total
mass contained in it, as we do here for the MW, and tracing its
history back to recombination and before.

\section{THE PAST: BACK TO RECOMBINATION (PHOTON DECOUPLING)}

It has become clear that the dominant galaxy formation processes
in our universe are bottom-up -- smaller galaxies formed first,
and then gradually agglomerated into larger galaxies and clusters
of galaxies. This is the principal basis for choosing the
worldlines of the matter that made up our Galaxy as the
cosmological domain of influence for the solar system, and then
focusing on its likely cosmological history all the way back to
the Planck era. This choice is strongly supported by the fact that
the local cosmological environment in which our Galaxy finds
itself is both noticeably underdense relative to nearby regions of
our local supercluster, hence it is not accreting matter from much
farther out,  and is surprisingly quiescent as a local peculiar
velocity field (within $5 h^{-1} Mpc$ around the MW) (see Klypin,
et al., 2003, and references therein).\footnote{Here and elsewhere
in this paper $h \equiv H_0/100$ km/sec/Mpc.} Despite this there
is continual accretion of material on smaller length scales,
including accretion of much smaller systems onto the MW (Cho 2008;
Ibata and Lewis 2008),  leading to what has been referred to as our
local ``cosmic web.'' On a much larger scale, of course, there is
Virgocentric infall. These discoveries
strongly support the bottom-up account of galaxy assembly from a
small local region around the galactic core of the MW.

Considering that all the matter presently constituting the MW was in the same
narrow tube of timelike geodesics all the way back to
recombination at $z_{rec} \approx 1100$, what was the length scale
embracing this bundle of world lines? The condition for the
dynamical importance of nothing larger than this bundle is
certainly fulfilled, since the effective pressure --and therefore
the sound speed -- of the matter in the universe is zero from
recombination/decoupling to the present, and the strengths of any
other possible long distance influences -- electromagnetic or
gravitational waves, or electric Weyl tensor components from
sources outside that region -- are insignificant compared to local
effects (Cox 2007). This is due to the large-scale almost-FLRW
character of our universe, and to the relative weakness and
$r^{-2}$ fall-off of these interactions. The effective domain of
influence for the MW as an evolving local system, then, is determined by the peculiar
velocities of the material in the local gravitational field of the
growing density fluctuations that eventually merge to form the MW.
This defines the sphere of matter which has had significant influence on
conditions here at the present time; in particular, it
characterizes the matter out of which the MW is composed.

Estimating the size of this domain of influence at $z_{rec}$ 
is relatively simple. At
the time of recombination, just when the anisotropies of the CMB
were imprinted, all the matter in the MW would have been in a
localized cluster of nearby density perturbations just slightly more
dense than the background at that time, by 1 part in a 100,000 (the
density perturbations, as we know from CMB data, were about $\delta
\approx 10^{-5}$ on a broad range of scales). If we take the total
mass of the MW to be $7 \times 10^{11} M_{\odot}$ within $100$ kpc
(Binney and Merrifield 1998) and assuming $\Omega_m = 0.3$,
$\Omega_{\Lambda} = 0.7$ and $H_0 = 70 $ km/sec/Mpc, a simple
continuity argument shows that the MW perturbations occupied a
volume of radius $1.46$ kpc at recombination, $z_{rec} = 1100$. This
is equivalent to a comoving length scale (relative to the present
time) of $d_{doi}(t_0,t_{rec}) = 1.6$ Mpc. 

Nothing on larger scales
has had a significant physical effect on conditions in our cosmic
neighborhood at the present time.\footnote{If the MW has 3 times the
amount of matter quoted here, as suggested by some references, e. g.
Peebles 1992, then the comoving size of the MW region at
recombination will be a little larger -- a radius of about $2.3$
Mpc.} The present comoving size of the particle horizon in an
Einstein de Sitter universe, which is essentially the size of the
visual horizon in that universe, is $d_{ph E_dS}(t_0) \simeq (1/3)
\times 10^{4}$ Mpc, so the effective domain of influence $d_{doi}$
at that time is very much less than this:
$d_{doi}(t_0,t_{rec})/d_{ph EdS}(t_0) \simeq 5 \times 10^{-4}$. In
an inflationary universe, the particle horizon is very much bigger
than the visual horizon (everything we can see), and the ratio is
enormously smaller. The change from null to timelike causation
radically redefines the domains of causal importance in cosmology. 

It is enlightening to ask a few simple questions about this
result. The first is: What size would such a cluster of
perturbations subtend on the LSS for a distant
cosmic observer who was trying to resolve it as a fluctuation in
her CMB data at the present cosmological time? The answer is that
it would subtend only about $0.6$ arcmin, or a spherical harmonic
index $l \approx 5370$, and considerably less than the thickness
of the LSS itself. The on-going Atacama Cosmology
Telescope (ACT) survey will probe this angular scale and smaller, but
a detailed analysis of decoupling dynamics would be needed to see if
traces of these inhomogeneities would remain in the CBR data.
 
But what are the conditions for such a perturbation to be present at
that time? Isn't it small enough to be wiped out by photon-diffusion
during decoupling? If the only form of matter present were baryons,
then, yes, a comoving perturbation of $1.6$ Mpc would indeed be
dissipated by that process (Kolb and Turner 1990, p. 355, equation
(9.97)). For an  FLRW background cosmology with no CDM and an $\Omega_b = 0.04$,
for instance, perturbations on co-moving length scales smaller than
about 30 Mpc would be dissipated by photon diffusion. However, if
the dominant matter in the universe is non-baryonic, as seems to be
the case -- there is about 6 or 7 times more CDM in
the universe than baryons -- the picture changes completely. Then
the CDM perturbations decoupled long before recombination and began
to grow. During the decoupling of the baryons from the photons they
would not be markedly affected by photon diffusion, and the baryons
will begin to fall rapidly into those CDM potential wells tracing
the CDM perturbations themselves. We shall return in the next
section to discuss the decoupling and evolution of these crucial CDM
fluctuations. The baryon dynamics after decoupling is purely
gravitational and thus dominated by DM.
 
A second question about the early growth and history of our MW
galaxy is: How large would the perturbations that eventually led to
the MW have been at recombination? How
big were the individual pre-MW size lumps that existed on the LSS?
\footnote{We refer here to perturbations \emph{like} those that led to
the MW. We cannot see the MW perturbations
themselves, no matter what size they were, as they were where our
past world line intersected the
LSS, nowhere near where our past light cone intersects it.}
There was undoubtedly a cluster of significantly smaller
perturbations, which were incorporated into the MW over time,
eventually attaining a comoving size of $1.6$ Mpc (and so a physical
size then of $1.4$ Kpc). A number of observational and theoretical
studies have indicated (Carlberg, et. al. 2000; Murali, et
al. 2002), for instance, that galaxies of the size of our MW have
grown by as much as factors of 2 to 3 by mergers and accretion since
$z = 1$ (about 7 Gyr for our cosmology). The amount of growth by
these two processes before then would have been even more
significant. Thus, the individual perturbations which combined to
constitute the MW were undoubtedly originally much smaller than the
$1.6$ Mpc figure, and so even more difficult to resolve in the CMB
anisotropy spectrum: they would subtend less than $0.3$ arcmin, or a
spherical harmonic index greater than $l \approx 10,740$. But we do
not know what the mass-spectrum of these individual perturbations
was.

\section{THE PAST: FROM INFLATION TO RECOMBINATION}
The question now is how far back did the domain of influence
remain so small in comoving terms. It would have remained small as
long as CDM dominated the dynamics, and CDM and baryons were
decoupled from the radiation. But how do we know that there were CDM
perturbations already growing before recombination allowed the
baryons to fall into them? And what are the conditions constraining
their size and determining their lower-end cutoff?

As regards the first question, we have nearly incontrovertible
evidence from primordial nuclear abundances together with velocity
dispersion measurements in clusters of galaxies, rotational curves
in spiral galaxies, etc., that there must be a much larger
percentage of matter in cold nonbaryonic particles than in baryons.
Furthermore, this is consistent with the result that it would be
impossible for the density fluctuations we see at last scattering
(in the CMB) to grow into the structures we observe now if only the
baryons were involved. There is not enough time. As we indicated
above, there would not have been perturbations on comoving scales
less than about $35$ Mpc (because of photon diffusion at
decoupling). 

Thus the standard understanding is that CDM perturbations are crucial
to structure formation in the
universe, essentially dominating and accelerating the growth of
baryonic perturbations after decoupling. It is the origin and growth
of these CDM perturbations ancestral to the MW and to other galaxies
that we need to focus upon. The smallest systems today which are dominated
by primordial dark matter are dwarf irregular galaxies, with mass of
the order of $2 \times 10^7 M_{\odot}$. These
would originate in the comoving range of, say $0.05$ to $1.0$ Mpc,
and would undergo mergers and accretion as perturbations -- and
later as galaxy haloes -- with the expansion and cooling of the
universe.
 
The answer to the second question depends on the epochs that occurred in
the past. Would not the region of importance have been determined by the 
speed of light before decoupling? No, because the universe was
opaque then: the mean free path for photons was very small. No
influences, except gravitational waves, could travel at the
speed of light, even when the universe was radiation dominated.
The way this worked out varied with epoch and with wavelength.

\subsection{The relevant epochs}
The relevant epochs before decoupling are as follows, in the order
in which they occurred:
\begin{itemize}
\item Before inflation (a quantum gravity domain?);
\item The inflation epoch (before reheating);
\item Reheating (the end of inflation) to CDM decoupling;
\item CDM decoupling to matter domination;
\item Matter domination to radiation decoupling (the LSS).
\end{itemize}
The outcome is different for different matter scales, because that
crucially affects the time of horizon exit (during inflation) and 
re-entry (after inflation). One should note that the ordering
of events given here depends on the relative energy densities of
matter, radiation, and dark matter: if they were different, the
relevant epochs might be different. So the analysis given here is
model dependent to that extent; it may differ with different numbers
for these densities. But it is in accord with the best present
estimates of how things are. 

We now look at each of these epochs in turn, from inflation on, for the 
physical scale of the MW region. The epoch before inflation is not
well understood, but also has a minimal effect on what occurs after
inflation, because the exponential expansion during inflation wipes
out memory of conditions before. In this sense the start of the 
history of he known part of the universe is during the inflationary
epoch. The preceding era is both unknown, and largely irrelevant
(given that it set up the conditions for inflation to occur).
 
A summary of the situation is given in Figure 1, which traces the
effective domain of influence for the MW region, and its matter
horizon, back to earlier and earlier times. These results are all
based on standard calculations, such those summarized in Dodelson
(2003).

\begin{figure}
\includegraphics[width=15cm]{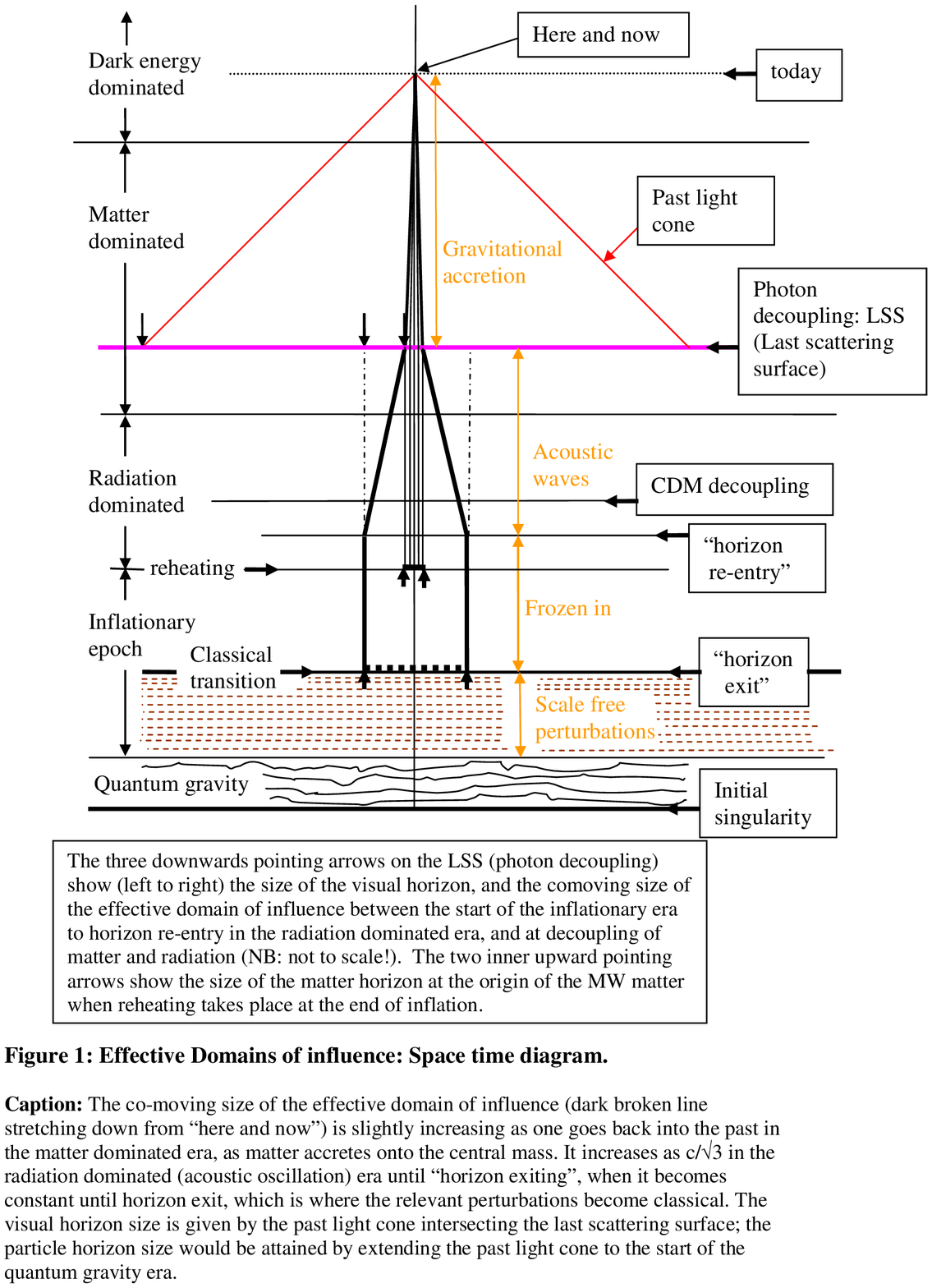}
\end{figure}

\subsection{Inflation and horizon re-entry}
To begin with, during inflation the MW perturbations were continuously
generated inside the effective horizon $H^{-1}$, but they would have
quickly left $H^{-1}$ during the inflationary period itself and become
``frozen in.''
 
Before the MW perturbations re-entered ``the horizon'' (the Hubble 
scale) at $t_{hor}$, very soon after reheating at the end of inflation,
there was no possibility of causal contact with other perturbations
through acoustic oscillations -- that is, at $t_{exit}$ they quickly
left ``the effective horizon'' $H^{-1}$ after their generation during
inflation. Before leaving, no correlations or causal connections that 
would persist into the classical regime were established. Remember that
before $t_{exit}$ the perturbations are purely quantum, and as such have
no realized classical status except as probability waves.\footnote{Once
the perturbations leave the horizon, they do, in fact, become classical,
or at least quasi-classical. See Linde 1990, Liddle and Lyth 2000, and 
Kiefer and Polarski 2008.} So there is no expansion of the effective 
causal domain due to acoustic waves or light-like waves earlier than
$t_{hor}$.

\subsection{From horizon re-entry to CDM decoupling} 
When the universe reheated at the end of inflation, not only would
the universe have been radiation dominated, but also there would
have been a certain period of time thereafter during which the
dominant dark-matter particles, as well as the baryons, would have
been coupled to the radiation and other matter in the universe by
interactions other than gravity. Under these conditions the
relativistic sound speed $c_s = c/\sqrt{3}$ set the maximum
limit of this dominant causal influence at that epoch. This determines
the effective horizon of communication at that time. Thus, from
horizon re-entry at $t_{hor}$, shortly after the end of inflation, 
until $t_{eq}$ -- when the universe became matter-dominated --
causal influences propagated and correlated the density perturbations
within a much larger region than given by the matter horizon, through 
the acoustic waves travelling at $c_s$. No influences, except for 
gravitational waves and neutrinos will travel over large distances
faster than $c_s$. Photons always have a velocity $c$, but in a dense
ionized plasma electromagnetic information will be diffusing outward
from a source at considerablly less than that because the photon mean
free path is so short.
 
Gravitational waves and some decoupled weakly interacting relativistic particles
(e. g., neutrinos) would have higher velocities than this, but would
not, in the $\Lambda$CDM cosmologies, have any significant effects on
perturbation size or growth. Though the MW CDM perturbations would
remain intact and grow between horizon re-entry just after
inflation/reheating and CDM decoupling, there would have been little
or no coalescence or merging -- they were still very tenuous, in the
linear regime, and dominated by and expanding with the Hubble flow.
However, because of the coupling of matter to radiation they were
subjected to coherent CDM acoustic oscillations, which propagated
through the medium at $c_s$. There will also be some matter diffusion,
but at a much slower speed: there will be very little spreading due to
this cause (the universe is expanding very fast then, and that will 
certainly dominate diffusion effects).
 
Since they are relatively small-scale, the MW CDM perturbations re-enter
$H^{-1}$ relatively early, at $t_{hor}$ during the radiation-dominated
era. While outside $H^{-1}$ after the end of inflation, they were ``frozen
in'' but still able to grow in size as $a^2(t) \propto t$, since the radiation
pressure was not effective over superhorizon distances (i. e., they were not
causally self-connected). Thus, in summary, as soon as each MW-size fluctuation
cluster -- and before that each of its components -- re-entered the effective
horizon at these very early times during the radiation-dominated era, it was in 
internal causal contact, and growing logarithmically. Before re-entering the 
effective horizon but after the end of inflation, it would not have been in
causal contact, but growing as $a^2(t) \propto t$. However, the matter horizon
extends virtually unchanged all the way back from decoupling at the LSS to
reheating at the end of inflation, when the scalar-field fluctuations generated
by inflation are transformed into matter-radiation density perturbations. This
is the origin of the matter that exists in the MW domain. 

\subsection{From CDM decoupling to matter domination}
Decoupling of the dark matter itself from the other particle species  
depends crucially on the particular dominant dark-matter particle, and
on its interactions with other particles. As long as its interaction
rate $\Gamma$ with other particle species is greater than $H$, thus
keeping it in equilibrium with the universe, it will be coupled. When
$\Gamma < H$ it effectively decouples from the system -- it evolves
separately from the other particle species (see Kolb and Turner 1990, p.
115, pp 119-130). Unfortunately, we know neither the identity nor the
distinguishing characteristics of these CDM particles. Thus, we
really cannot determine when this decoupling occurs. Since the
perturbations are CDM, though, we do know that the particles must
have decoupled when the they were non-relativistic -- that is when
$m_{CDM} >> T_{d}$, where $T_d$ is the temperature at which
$\Gamma_{CDM} \leq H$. This means that, even though they decoupled
long before $z_{eq}$, they are massive enough so that they have
relatively small random thermal velocities. This is characteristic
of all CDM.
 
However, once the CDM particles decouple, their perturbations are
subject to damping by free-streaming. CDM particles can stream out
of the overdensities, causing them to evanesce. This can only
happen for short length-scale inhomogeneities. Kolb and Turner
(1990, pp. 352-353) calculate the general free-streaming comoving
length scale for such particles as
\begin{equation}
\lambda_{FS} = 0.2 Mpc (m_X/keV)^{-1} (T_X/T)[ln(t_{eq}/t_{nr}) +
2], 
\end{equation}
 where $m_X$ is the mass of the CDM particle in
keV, $T_X$ is the temperature of the CDM particles, $T$ is the
temperature of the photons, $t_{nr}$ is the time at which they become
non-relativistic, and $t_{eq}$ is the time of radiation-matter
equality. Generally speaking for CDM particle candidates $T_X$
will be significantly less than T, and $t_{nr}$ will be
significantly less than $t_{eq}$.
 
Thus, until we know more about
the dominant CDM candidate, we cannot reliably establish the
length-scale below which our CDM perturbations will be
significantly damped by free-streaming. It is very likely, though,
that the primordial CDM perturbations of the size that triggered
the early formation of what became the MW remained intact.
Furthermore, they certainly gradually accumulated substantially
more mass by accretion and through mergers with other nearby
perturbations. We shall presume this in the rest of this paper. If
they did not remain intact, then it is not possible to explain the
plethora of structure we see on small (galaxy-sized) scales, given
the evidence for bottom-up galaxy formation. A complementary
top-down structure formation process for such small-scale, MW-size
perturbations is not supported by the evidence. From equation (1),
we can see that it is not hard to account for persisting CDM
structures on scales $\approx 1$ Mpc, as long as $m_X$ is not much
less that $1$ keV.

After CDM decoupling, but before photon decoupling, the baryons were
still tightly coupled to radiation (photons to electrons via Thomson
scattering, protons to electrons via the Coulomb interaction). Thus the
baryons cannot collapse because of radiation pressure, whereas the CDM
is not so constrained. The complication here is that between CDM decoupling at $t_{CDM}$
and matter-radiation equality later at $t_{eq}$, the universe was dominated
by radiation coupled with the baryons. Though the CDM was decoupled from that
mixture, it was dominated gravitationally by the radiation and therefore forced to move
through gravitational coupling with baryon-radiation acoustic oscillations.
Once the universe became matter dominated (at $t_{eq}$), the CDM gravitationally
dominated the baryon-radiation fluid, and the decoupled CDM perturbations
modulated the baryon-radiation acoustic oscillations.

In the period right after $t_{CDM}$ -- certainly while radiation
was still in control (until $t_{eq}$) -- CDM perturbations which
had already re-entered the effective dynamic horizon $H^{-1}$ did not
grow significantly (see, for instance, the detailed treatment of 
Dodelson 2003, especially Section 7.3). Fluctuations began to grow
approximately logarthmically, with a slightly more enhanced rate of
growth as the matter-dominated regime was approached, at $t_{eq}$. 
This early relative suppression of small length scale perturbations
has been detected in the mass spectrum. It is possible that before
CDM decoupling diffusion altered the perturbations, but we have not
taken it into consideration, because the effects would not be large.

\subsection{From matter domination to photon decoupling}
For our universe, with $\Omega_{m0} = 0.3$, the redshift $z_{eq}$ of 
radiation-matter equality would be roughly at $1 + z_{eq} \approx
3500$. This is considerably before recombination at $z_{rec} \approx
1100$. The age of our universe at $t_{eq}$ when the matter and radiation
densities are equal, assuming our universe has $\Omega_{\Lambda 0} 
\approx 0.7$, $\Omega_{tot} = \Omega_m + \Omega_{\Lambda} = 1$, was
$t_{eq} \approx 8.9 \times 10^3$ years. 
 
If our universe were not expanding, causal influences could
effectively propagate a distance $c_s t_{eq} = (c/\sqrt{3}) 8.87
\times 10^3 \; years = 5.12 \times 10^3 \; lightyears = 1.57
\times 10^{-3} \; Mpc$ by $t = t_{eq}$.\footnote {Here we are
assuming that there is no substantial contribution from the
extremely brief period between the Big Bang and the end of
inflation. Since the sound speed for a scalar field will be $c$,
anyhow, that will not make any difference.} However, our universe
{\it is} expanding, and so the effective domain of influence
$d_{s}(t)$ at $t = t_{eq}$ will really be given by: 
\begin{equation} d_s(t_{eq}) =
\frac{a(t_{eq})}{\sqrt{3}} \int_0^{t_{eq}} dt'/a(t').
\end{equation}
Using the fact that in the
radiation-dominated era $a(t) \propto t^{1/2}$ (even if
$\Omega_{\Lambda 0} = .7$, $\Omega_{\Lambda eq}$ will be extremely
small, as can be easily shown), we find that
\begin{equation}
d_s(t_{eq}) = 2 c
t_{eq}/\sqrt{3} = 3.14 \times 10^{-3} \; Mpc.
\end{equation}
As a co-moving distance, relative to distances at our redshift
now, our result is $d_s(t_{eq})_0 = 11.08 \; Mpc.$ Comparing this
with the co-moving length scale of $1.6 \; Mpc$ for our MW cluster
of perturbations at $t_{rec}$, which is later than $t_{eq}$, we
confirm that MW-size fluctuation clusters were already well within
the cosmological causal limit at $t_{eq}$. It was also well within
the Hubble scale $H^{-1}$, which is often referred to as ``the
horizon'' in terms of dynamical dominance. 

Once the MW perturbations enter the matter-dominated regime ($t
> t_{eq}$), they grow as $a(t) \propto t^{2/3}$, much more rapidly
than logarithmically (we are assuming here a flat universe). There
are very slight modifications to this
basic picture due to the presence of baryons and dark energy. This
growth, of course, presumes that that MW CDM perturbations are
stable against free-streaming, as indicated above. This is why
confirmation that the mass of the dominant CDM particle is large --
to explain the survival and early growth of smaller CDM fluctuations
-- is important to verify. After recombination, as we have already
mentioned, the baryons are then free of the radiation, and can
follow these CDM perturbations, falling into their potential wells.

But what about the remaining baryon-acoustic oscillations before
recombination and photon-baryon decoupling when the universe is
dominated by CDM? During that period long-range correlations can be
established via those sound waves -- but now modulated, as we have
just said, by the presence of the developing CDM potential wells.
Since all the growth in those perturbations and their gravitational
influence is still in the linear regime, those baryon-acoustic
oscillations will not be trapped in those perturbations and will
communicate across them. Thus, we need to extend the growth of the
effective domain of influence forward to the $t_{rec}$, the time of
photon-baryon decoupling and recombination. This causal limit, which
has often be referred to as the ``sound horizon'' or the ``acoustic
horizon'' has been well recognized and discussed extensively in the
literature since at least 1970 (see Peebles and Yu 1970; Sunyaev and
Zeldovich 1970; Eisenstein and Hu 1998; Dodelson 2003; and
Desjacques 2008, and references therein). It is represented in
Figure 1 by the diagonal broadening of the domain of influence from
the matter horizon at $t_{rec}$, not just $t_{eq}$, all the way back
to $t_{hor}$. This means that we should find density correlations in
the large-scale structure on that scale -- and indeed we do, as
enhancements in the galaxy and cluster-of-galaxy correlation
functions at about 105 $h^{-1}$ Mpc (Eisenstein, et al. 2005; Cole,
et al. 2005; Estrada, et al. 2008; Desjacques 2008). This
corresponds, as is very well known also, to the the position of the
first peak in the temperature anisotropy measurements of the CMB
(see, for instance Hu, et al. 2001 and Doran and Lilley 2002, and
references therein).

Theoretically, how is that scale explained? Equation (3) shows that
$d_s(t_{eq}) = 2ct_{eq}/\sqrt{3}$, where we are assuming that
horizon re-entry for the MW perturbation occurs just after
inflation ends at $t \approx 0$, relative to $t_{eq}$. To this has
to be added the extra distance due the acoustic wave propagation still 
being effective during the early matter-dominated epoch from $t_{eq}$ to
$t_{rec}$. This will be
\begin{equation}
d_s(t_d) = \sqrt{3}c(t_{rec} - t_{rec}^{2/3}t_{eq}^{1/3}),
\end{equation}
where $t_d \equiv t_{rec}-t_{eq}$. Then the distance over which
density correlations were induced by the acoustic oscillations
between horizon entry and recombination is simply
\begin{equation}
d_s(t_{rec}) = d_s(t_{eq}) + d_s(t_d). 
\end{equation}
 This distance defines a correlation length at horizon re-entry for the
perturbations immediately surrounding the MW perturbations -- and
constitutes the effective domain of influence for the MW
perturbations at that time. It gives the limit out to which the
dominant causal influences have significantly impacted and connected that region
and the objects it contains. It is much larger than the matter
horizon, which specifies the limit from which material has been
permanently added to our region.

In principle and in practice, as we just
mentioned above, this effective domain of influence is observable
now in intermediate-scale density correlations in the nearby
universe -- as ``the baryon acoustic oscillations'' recently
detected (Eisenstein, et al. 2005). $d_s(t_{eq}) \approx 3.14 \times 10^{-3}
Mpc$, which translates into a comoving distance
(relative to the present) of $11.08 Mpc$. Using the simple
relationship between $t_{rec}$ and $z_{rec}$ (see Kolb and Turner
1990, p. 80), we find from the relationship above that $d_s(t_d) =
9.13 \times 10^{-2}$ Mpc, or 100.4 Mpc, comoving. Thus, the comoving
scale of the realistic causal or sound horizon $d_s(t_{rec})$ is
about 111 Mpc. This corresponds to a spherical harmonic index of $l
\approx 80$, or about 38 arcmin., which would be an easily observable
angular scale in CMB data.

\subsection{The effective domain of influence}

Thus, in the standard $\Lambda$CDM universe -- or in any one not too 
different from it --  the topography of local causal domains significant for our
cosmic region illustrated in Figure 1 involves the narrow core of the
matter horizon extending from the MW all the way back down to the
generation of its components in the reheating at the end of
inflation, with a comoving size of about 2 Mpc. Shrouding this matter
horizon is the effective domain of
influence, expanding out to approximately 110 Mpc through the
effects of CDM-radiation and baryon-radiation acoustic waves between
$t_{hor}$ and $t_{rec}$. This more extensive domain which
characterizes the causal history of our locale before $t_{rec}$ does
not add material to the MW perturbations themselves nor significantly
alter the world-lines along which they flow, but rather correlates
them them with other clusters of perturbations due to the
causal influence of the acoustic oscillations on the CDM and the baryons
before recombination and baryon decoupling. 

The effective domain of influence is much smaller than both the particle
horizon and the visual horizon, firstly because its broadening out
as we go back into the past from the LSS starts from the scale of the 
matter horizon at the LSS, not from the scale of our past light cone;
and secondly, because it broadens out at an effective speed of 
$c/\sqrt{3}$, rather than $c$. Furthermore it effectively stablizes
at horizon re-entry, remaining the same size until the start of the
classical domain, where we can begin to sensibly follow its evolution.

\section{Present Domain of Influence and Future Matter Horizon}
So far in this paper we have, from the perspective of the past
history of the universe, considered the MW as our local dynamical
region, and its extent as our present matter horizon. However, as we
have emphasized, the MW itself is continuing to grow by accreting
material, including smaller systems, from its environment, and is
gravitationally interacting with nearby and more distant galaxies
and clusters -- most clearly, the members of the Local Group, and
the Virgo cluster. Taking these present interactions with other
systems into account, as well as those in the distant future, can we
physically and quantitatively characterize, or define, the growth of
the matter horizon and the effective domain of dependence as the
universe evolves into the future? This would give
us the dynamical limit of our cosmic locale. 

We have already briefly mentioned several possible definitions. As we have
emphasized, as long as our universe continues to be dominated by $\Lambda$CDM, with
no extreme exotic surprises, which  we have been assuming, there are vast regions of it 
which will never alter, or significantly influence, our own locality --
whose world lines will never approach or intersect our own world line. There
are regions at a sufficient distance from us -- but not at an
extraordinary distance from us -- which are now expanding away from us. And
there are others, somewhat further out, which will never collapse towards us --
no matter how far into the future we go. 

Motivated by these general observations about the relationship
between our locality and the universe, we define both a present
domain of influence and a future matter horizon. This is related to the
idea of a finite infinity surrounding us -- see Ellis (1984) and Cox
(2007). The way they are related to each other is indicated in Figure
2. 

\begin{figure}
\includegraphics[width=15cm]{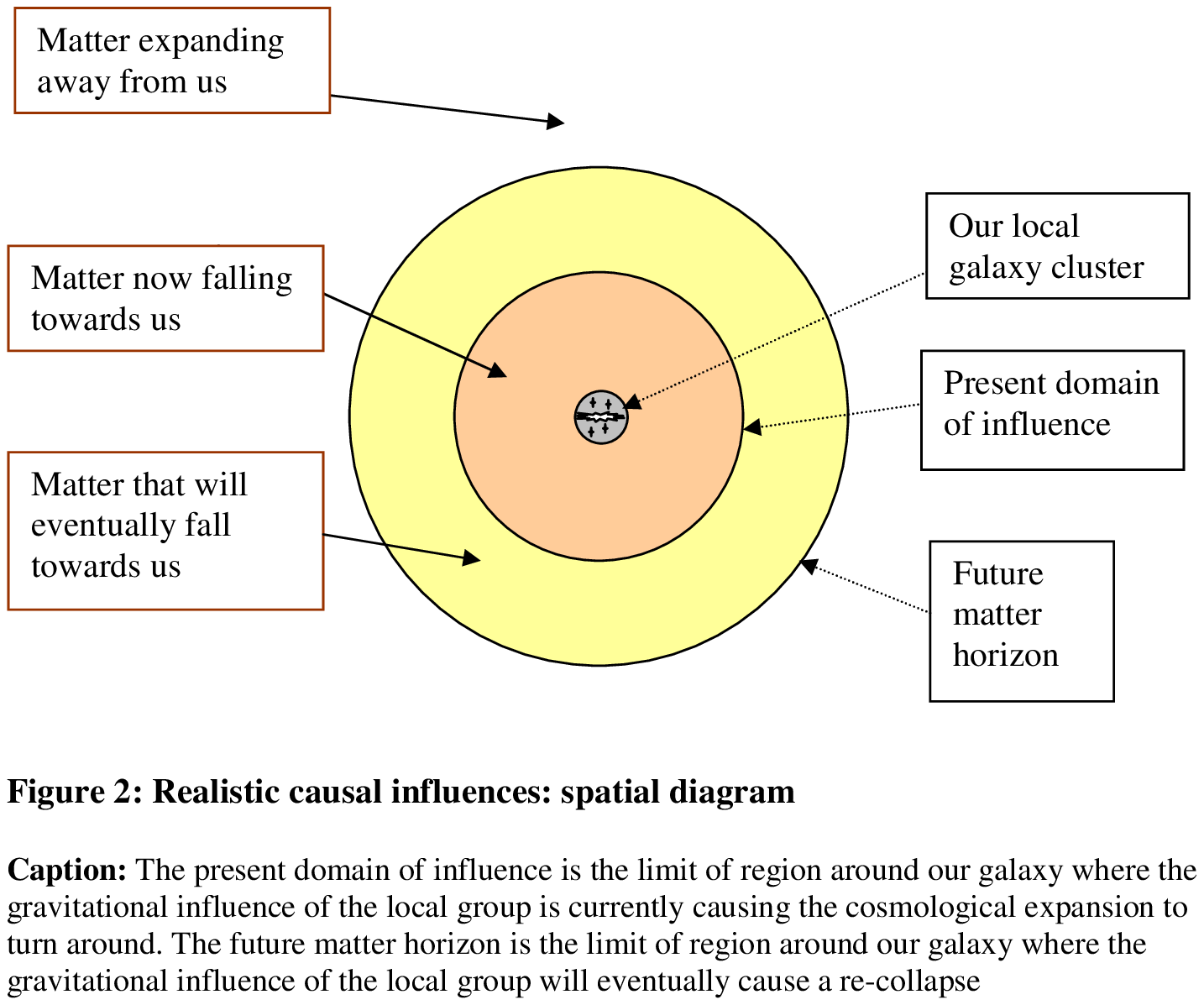}
\end{figure}

\subsection{ The future matter horizon}
The {\it future
matter horizon} is, ideally, the spherical shell \footnote{It is
very clear that the MW is {\it not} at the center of our dynamical
local domain -- with Virgo-centric infall and the motion of our
neighboring clusters of galaxies towards the Great Attractor.
However, here we simplify our analysis to spherical symmetry to
formulate and examine the suitability and implications of the basic
concepts.} at the radial distance $r_{oh}$ from any point $q_0$ --
e. g. our own location -- beyond which no world-lines will ever
converge towards $q_0$, beyond which nothing will ever collapse
towards us. 

In working out the criterion for the future matter horizon $r_{oh}$
we can apply the approach Padmanabhan (1993, 275-276) uses to  work
out in Newtonian approximation the condition for the eventual
collapse of an overdense region in an expanding universe. The
condition itself is on $\bar{\delta}_i$, which is the average
overdensity of the region out to a radius $r_i$ at a particular time
$t_i$ relative to the background cosmic density at that time. This
key parameter is given by
\begin{equation}
 \bar{\delta}_i = \frac{3}{r_i^3}\int_0^{r_i} \delta_i(r) r^2 dr,
\end{equation}
 where $\delta_i(r)$ is the local density contrast.
Essentially, as we shall see, the future matter horizon $r_{oh}$
is the value of $r_i$ in equation (6) -- in both the denominator
and in the upper limit of the integral --  which gives the value
of $\bar{\delta}_i$ which is just low enough so worldlines at
$r_i$ will never converge towards us. It is the boundary between
the region around us which is collapsing or will eventually
collapse, and the region beyond, which will never approach us and
continues to expand forever. In implementing this criterion we
first must determine what value of $\bar{\delta}_i =
\bar{\delta}_{oh}$ insures that, and then from observational
measurements of $\delta_i(r)$ determine at what value $r_i =
r_{oh}$ that value is reached. Padmanabhan (1993) worked out the
first part of this problem in general for a $\Lambda = 0$ FLRW
universe. Here
we shall do the calculation for our case -- where $\Lambda \neq 0$. 

Assuming spherical symmetry and using the Newtonian approximation, we have the
energy $E$ for a given overdense region in an FLRW universe
\begin{equation}
E = \frac{1}{2}(\frac{dr_i}{dt})^2 -
\frac{GM}{r_i}+\frac{\Lambda}{3}r_i^2,       
\end{equation}
where $r_i$ is the radius of the region, $M$ its mass, and
$\Lambda$ is the vacuum energy (cosmological constant). The
condition for eventual collapse is
obviously $E < 0$. 

We cannot use Padmanabhan's simplest calculation, since it assumes that $\bar{\delta}_i$
is small enough so that the boundary of the region at $r_i$ is still expanding with the
Hubble flow. This will certainly not be true in our case, with $\Lambda \neq 0$, as we
shall see. Thus, we use his generalization (outlined in Padmanabhan 1993, p. 320,
Exer. 8.1) to obtain
\begin{equation}
\bar{\delta}_{oh} \geq (1+\frac{v_{oh}}{H_0
r_{oh}})^2\Omega_m^{-1} + \frac{2\Omega_{\Lambda 0}}{\Omega_{m0}}
- 1, 
\end{equation}
 where we are considering the situation now ($t =
t_0$) at $r_{oh}$, and where $v_{oh} < 0$ is the peculiar (radial)
velocity there due to the local gravitational
potential within $r_{oh}$. 

Further, we know from the detailed work of an number of
researchers (see Lahav, {\it et al.} 1991, and references therein)
that $v_i$ is related to the $\bar{\delta}_i$ within the region by
\begin{equation}
v_i = \frac{1}{3}H_0r_i f \bar{\delta}_i, 
\end{equation}
where $f$ is a function of $\Omega_m$ and $\Omega_{\Lambda}$. For
our time now it is given (Lahav, {\it et al.} 1991) by
\begin{equation}
f_0
\approx \Omega_m^{0.6} + \frac{1}{70}\Omega_{\Lambda}(1 +
\frac{1}{2}\Omega_m). 
\end{equation}
Thus, combining equations (8), (9) and (10), we solve a quadratic
equation to obtain, for our values of $\Omega_{m0}$ and
$\Omega_{\Lambda 0}$,
\begin{equation}
 \bar{\delta}_{oh} \geq 4.0. 
\end{equation}
This means that for all $r_i < r_{oh}$, such that
$\bar{\delta}_{oh}$ is given by equation (8), or in our case by
equation (11), the world lines will eventually undergo collapse,
even if they are still expanding now.

With the result (9), if we also know $\delta_i(r)$ throughout the
region bounded by $r_{oh}$, we can use equation (6) to find the
value of $r_{oh}$ itself. Unfortunately, we do know $\delta_i(r)$,
or its non-spherically symmetric generalization, far enough out to
determine $r_{oh}$ reliably. But this should
certainly be possible to estimate in the near future. 

One clear consequence of this simple calculation is that $r_{oh}$ in
our observable universe will {\it not} be the boundary beyond which
the matter flow lines are moving with the Hubble flow. Furthermore,
it is also clear that even for a range of values of $r > r_{oh}$,
the geometry is not yet perturbed FLRW. That boundary is still
farther out, at $r_b$ such that $\bar{\delta}_b < 1$. This is also
obvious from the peculiar velocity at $r_{oh}$, $v_{oh} = -0.67
v_{red}$, where $v_{red} \approx H_0 r_{oh}$ is the ``redshift
velocity'' at $r_{oh}$. Thus, outside $r_{oh}$ there will be a
fairly large region which, though influenced by the Hubble expansion
and fated to expand forever, is at present significantly influenced
by the peculiar velocity due to $\bar{\delta}_{oh}$.

\subsection{The Present Domain of Influence}
The {\it present domain of influence} $r_{ih}$ is
defined by the surface at the largest distance from our position
where the matter flow lines are moving towards us (collapsing) at
our time now, $t = t_N$. Beyond that radius they are still dominated
by cosmic expansion. It is the radial distance from $q_0$ beyond
which expansion from us is still occurring, and inside of which
there is overall collapse of matter world lines towards us. We
define $r_{ih}$ both because it gives a clearer dynamical definition
of the actual local domain of influence at any given time and
because it can be defined in the case of a closed universe, whereas
$r_{oh}$ cannot be.

As we have already indicated, the boundary of the present domain of
influence $r_{ih}$ is the distance now to the shell of material
surrounding $q_0$ which is just about to begin collapsing. This is
the extent of the region centered on $q_0$ in which world lines are
now converging. Outside $r_{ih}$ they are presently influenced by
the cosmic expansion, and still diverging. The matter in this
region, obviously, will eventually affect our locale, and will do so
sooner than that at radii between $r_{ih}$ and $r_{oh}$. The
criterion for determining $r_{ih}$ is quite simple, and somewhat
similar conceptually to that for determining $r_{oh}$. Essentially,
at $r_{ih}$ the peculiar radial velocity $v_i$, which is induced by
the overdensity of matter $\delta_j (r)$ within the region, exactly
cancels the velocity $H_0 r_{ih}$ due to the cosmic expansion of the
universe. Here $r_{ih} = x_{ih}R_0$ is the actual physical distance
to the inner realistic horizon, where $x_{ih}$ is the comoving
distance to it and $R_0$ is the scale factor of the universe.  Thus,
at $r_{ih}$ we have
\begin{equation}
 H_0 r_{ih} = v_{ih}. 
 \end{equation}

Applying equation (9) above, we find that
\begin{equation}
\bar{\delta}_{ih} = 3f^{-1}, 
\end{equation}
 where, as before,
$\bar{\delta}_{ih}$ is given by equation (6) with $r_i = r_{ih}$.
With again $\Omega_m = 0.3$ and $\Omega_{\Lambda} = 0.7$ and using
equation (10), equation (13) gives
\begin{equation}
\bar{\delta}_{ih} = 6.1. 
\end{equation}
At smaller radii, where the world lines are converging,
 $\bar{\delta}_i$ will be larger than
this, and at larger radii, where the shells will be feeling the
cosmic expansion to some extent at least, it will be smaller.
Obviously, this result is consistent with our result for $r_oh$,
since $\bar{\delta}_{ih} > \bar{\delta}_{oh}$, as it must be, for
$r_{ih} < r_{oh}$. Once again, of course, in order to determine
$r_{ih}$ itself from our calculation of $\bar{\delta}_{ih}$, we
need to to know the local density contrast function $\delta_i(r)$
within that region
and solve equation (6). 

In principle, $r_{ih}$ could be observationally determined, by finding
the boundary where the total redshift (cosmological + peculiar-velocity) of
relatively nearby galaxies and clusters of galaxies is zero. This would enable
the direct
determination of $r_{ih}$ at the look-back time corresponding to that redshift.
$r_{ih}$ now, which is not directly determinable by observation, could then
be extrapolated from that result. Because of all the local
motions within the various sub-systems, the first step in this program would
undoubtedly require a great deal of careful observational and statistical work. \\

\section{CONCLUSION: THE DOMAINS OF INFLUENCE}
The effective domains of influence defined here  -- the matter horizon, the
effective causal horizon, the present domain of influence, and the
future matter horizon -- are obvious, important, but often
unarticulated, characteristics of our universe. The part of space-time that
significantly influences us is a very small part of our causal past,
as clearly shown in Figure 1. On intermediate and
large scales, the observable universe is so well modelled by
linearized gravitational effects, and the galaxies and clusters of
galaxies are so spread out, that conditions at even moderate
distances do not interfere with the local physics of the solar
system or of our Galaxy. However, paradoxically, this is because conditions
are set up in the universe so that isolated systems exist and function more
or less independently of distant regions. This need not have been true.

This might be described as an effect of the
geometry of our universe being so close to FLRW on large scales.
Interestingly enough, we could say from a
complementary perspective that this property {\it is} highly
relevant to local physics, because it insures that relatively
isolated systems like ours are able exist and evolve on their own.
Cosmology establishes the overall context within which that becomes
possible! And certainly, on the standard understanding, inflation was
crucial to insuring that our universe was homogeneous and diffuse
enough to manifest this localization property. 

To repeat the cautionary note we started with: there are other
exotic possibilities for the nature of matter in the universe 
than those assumed here. For example, we do not know for certain
that the dark energy does not have a large velocity relative to 
the baryon rest frame. A relativistic ``wind'' of dark energy 
could significantly alter the ``horizons'' discussed in this 
article (Maroto 2006, Jimenez and Maroto 2008). However, this is
not the standard view; if such physics were to eventually turn out
to be the case, what is presented here would have to be modified
accordingly. Given the standard context, the signficant causal
domains for our local region are as outlined in this paper and
summarized in Figure 1.

Two indirectly related issues are connected with this relative
insensitivity of local physics to other locales at cosmic
distances, The first is that we can only measure the integrated
effect of anisotropies on the velocities of matter in our region.
This is true also of local tidal
effects due to large lumps of matter at large distances from us.
In these cases the influence of each shell of material at large
distances will cancel -- unless the mass distribution is very
anisotropic.

Secondly, gravitational influences {\it are} felt at
large distances, even outside the conventional visual, particle or
event horizons, via the ``Coulomb'' gravitational interaction
(Ellis and Sciama 1972). Gravity  is a long-range force, and N-body
dynamics is chaotic -- thus the influence of the precise distribution
of matter in large-scale structure is expected to be significant.
It is exactly for this reason that numerical simulations of clusters
must include a much larger volume that the cluster itself to be
``realistic.'' Because they are closely associated with
the mass-energy-momentum conservation equations, and therefore
with the constraint equations of general relativity (those which
must be fulfilled on each space-like slice of space-time), these
effects are instantaneous. That is why they can be felt beyond
horizons. However, they cannot transmit new information: the
gravitational influence of a large distant system is unchanging,
because its mass-energy must be conserved. Because the external field
is almost isotropic, these effects wil not be signficant on 
larger scales than those associated with the Baryon Acoustic Oscillations,
and they do not change our conclusions. 

These two features of
gravity reinforce the very special large-scale character of our
universe, and the protection is offers to the dynamical integrity
of local regions.

In summary:

 1.) Since the time of decoupling of matter and radiation,
 the domain that really matters as far as local physics is concerned
  is the cluster of linear perturbations which became the MW, with co-moving size
  (relative to our present time) between 1.5 and 2.3 Mpc. This is much smaller than any
discernible small-scale anisotropy in the CMB. In fact, a
perturbation of that size would be unobservable in principle
through CMB observations, because it is of a length scale smaller
than the thickness of the last-scattering surface itself. Thus, we
cannot detect or study perturbations of this scale in the CMB.

2.) Once the CDM and the baryons decouple from the radiation during
the radiation-dominated and the early matter-dominated eras, the
local physics is dominated by the CDM matter flow along nearby
world-lines, rather than by electromagnetic or acoustic waves. This
is primarily due to the near homogeneity of the cosmic background,
and the lack of direct influence of conditions at even moderate
distances from a given locale. This becomes an even more dominant
feature after
radiation-matter equality, and later after recombination. \\

We thank Roy Maartens for comments and discussions, and a 
referee for useful comments. \\

\noindent
{\Large \bf References} \\

\noindent
Carlberg, R. G., et al. 2000. Ap. J. 532, L1. \\

\noindent
Cho, A. 2008. Science 319, 47. \\

\noindent
Cole, S., et al. 2005. Mon. Not. R. Astron. Soc. 362, 505. \\

\noindent Cox, D.P.G. 2007. Gen. Rel. Grav. 39, 87.\\

\noindent
Desjacques, V. 2008. arXiv: astro-ph/0806.0007v3. \\

\noindent
Dodelson, Scott 2003. {\it Modern Cosmology}, Academic Press, 440 pp. \\

\noindent
Doran, M. and Lilley, M. 2002. Mon. Not. R. Astron. Soc. 330, 965. \\

\noindent
Eisenstein, D. J. and Hu. W. 1998. Astrophy. J. 496, 605. \\

\noindent
Eisenstein, D. J., et al. 2005. Astrophys. J. 633, 560. \\

\noindent  Ehlers, J.,  Prasanna, A. R., and  Breuer. R. A., 1987.
Class. Quantum Grav. 4, 253.\\

\noindent Ellis, G. F. R. 1984. In \emph{General Relativity and Gravitation},
 Ed B Bertotti et al, Dordrecht: Reidel, 215.\\

\noindent Ellis, G. F. R. \& Sciama, D. W. 1972. ``Global and
Non-Global Problems in Cosmology,'' in {\it General Relativity:
Papers in Honour of J. L. Synge},
L. O'Raifeartaigh, editor, Oxford, Clarendon Press, pp. 35-59. \\

\noindent
Ellis, G. F. R. and Stoeger, W. R. 1988. Class. and Quantum Grav., 5, 207. \\

\noindent
Estrada, J., Sefusatti, E., and Frieman, J. A. 2008. arXiv: astro-ph/
0801.3485. \\

\noindent Hawking, S. W. and Ellis, G. F. R. 1973. {\it The Large
Scale Structure of Space Time}. Cambridge University Press.\\

\noindent
Hu, W., Fukugita, M., Zaldarriaga, M., and Tegmark, M. 2001. Astrophys.
J. 549, 669. \\

\noindent
Ibata, R. A. \& Lewis, R. 2008. Science 319, 50. \\

\noindent
Jimenez, J. B. and Maroto, A. L. 2008. arXiv:0811.3606v3. \\

\noindent
Kiefer, C. and Polarski, D. 2008. arXiv: astro-ph/0810.0087v1. \\

\noindent
Kolb, Edward W., and Turner, Michael S. 1990. {\it The Early Universe},
Addison-Wesley Publishing Co., 547 pp. \\

\noindent
Klypin, A., Hoffman, Y., Kravtsov, A. V., and Gottl\"{o}ber, S.
2003. Ap. J. 596, 19. \\

\noindent
Lahav, O., Lilje, P. B., Primack, J. R., \& Rees, M. J. 1991. MNRAS 251,
128. \\

\noindent
Liddle, A. R. and Lyth, D. H. 2000. {\it Cosmological Inflation and Large-Scale
Structure}, Cambridge, Cambridge University Press, pp. 177-185. \\

\noindent
Linde, Andrei 1990. {\it Particle Physics and Inflationary Cosmology},
4th Printing (1993), Harwood Academic Publishers, Chur, Switzerland. 362 pp. \\

\noindent
Maroto, A. L. 2006. JCAP, 0605.015. \\

\noindent
Murali, Chigurupati, et al. 2002. Ap. J. 571, 1. \\

\noindent
Padmanabhan, T. 1993. {\it Structure Formation in the Universe}, Cambridge
University Press, 483 pp. \\

\noindent Peebles, P. J. E. 1993. {\it Principles of Physical
Cosmology},
Princeton University Press, 718 pp. \\

\noindent
Peebles, P. J. E. and Yu, J. T. 1970. Astrophy. J. 162, 815. \\

\noindent Rindler, W. (1956). {\it Mon Not Roy Ast Soc} \textbf{116}, 662.\\

\noindent
Sunyaev, R. and Zeldovich, Ya.-B. 1970. Astrophys. J. Suppl. Ser. 7, 3. \\

\noindent Tipler, F J , Clarke, C J S, and Ellis, G F R (1980):
``Singularities and Horizons: a review article". In \emph{General
Relativity and Gravitation: One Hundred years after the birth of
Albert Einstein}, Vol. 2, Ed. A. Held (Plenum Press), 97-206.

\end{document}